\documentclass[preprint]{aastex631}

\received{DRAFT 2024 October 24}
\graphicspath{{./}{figures/}}

\begin{document}

\title{Continuous Fragmentation of Comet 157P/Tritton}

\correspondingauthor{Jane Luu}
\email{jane.luu@geo.uio.no}

\author[0000-0002-0786-7307]{Jane Luu}
\affiliation{PHAB, Department of Geosciences, University of Oslo \\
Institute of Theoretical Astrophysics, University of Oslo \\
P.O. Box 1047, Blindern, NO-0316 Oslo, Norway}

\author{David Jewitt}
\affiliation{Department of Earth, Planetary and Space Sciences, UCLA \\
595 Charles Young Drive East, 
Los Angeles, CA 90095-1567} 





\begin{abstract}

We observed the split comet 157P/Tritton in October - November 2022 and January 2024 with the Nordic Optical Telescope (NOT).  Our observations show that the splitting continued during the entire observing campaign. Fragmentation was associated with outbursts, consistent with the action of outgassing torques that spun up the nucleus and its fragments to the point of rotational instability.  The outburst-fragmentation events can lead to a runaway process where the increasing spin rate, driven by outgassing torques, results in repeated mass loss, until the sublimating body completely disintegrates. 

\end{abstract}



\section{Introduction} \label{sec:intro}

The orbit of short period  comet 157P/Tritton  has semimajor axis $a$ = 3.368 au (orbital period $P$ = 6.18 years), eccentricity $e$ = 0.627 and inclination $i$ = 11.2\degr.  The comet has an eventful observation history.   Following its discovery in 1978, 157P was missed at several perihelion returns and was subsequently lost until 2003, when it was mistaken for other objects;  it finally secured a reliable orbit determination in 2003.  Most recently it passed within 0.265 au of Jupiter in 2020 \citep{sekanina2023}, then reached perihelion at heliocentric distance $r_H$ = 1.572 au on UT 2022 Sep 09. Observations of 157P from three different observatories reported a companion during UT 2022 August 21 - September 2 and September 18 - 28 (Minor Planet Electronic Circular 2022-T23); the companion was subsequently designated “157-B” (Minor Planet Center 2022).  However, upon closer inspection, \citet{sekanina2023} found that those observations were of two distinct companions which faded and brightened in such a fashion  that they were mistaken for a single object.  

In this paper we present optical observations of 157P in which we detected not two, but multiple companions. Like the comet itself, the companions changed their appearance dramatically as the comet receded from the Sun.

\section{Observations} \label{sec:observations}

We obtained observations of 157P/Tritton with the 2.5m diameter Nordic Optical Telescope (NOT), located on La Palma, the Canary Islands; the nights of observation spanned 2022 October - November, followed by the night of 2024 January 31.  The observations made use of the Andalucia Faint Object Spectrograph and Camera (ALFOSC) optical camera, equipped with an e2v Technologies $2048 \times 2064$ pixel charge-coupled device (CCD). The camera has pixel scale 0.214\arcsec/pixel, resulting in a vignette-limited field of view of approximately 6.5\arcmin$\times$ 6.5\arcmin.  All observations were made in the broadband Bessel R filter (central wavelength $\lambda_c = 6500$ \AA, full-width at half maximum (FWHM) 1300 \AA).  

\begin{deluxetable*}{lcccc}
\tablecaption{Journal of Observations \label{tab:ObservationsTable}}
\tablewidth{0pt}
\tablehead{
\colhead{UT Date} & \colhead{$r_H$} & \colhead{$\Delta$} & \colhead{$\phi$} & \colhead{Exposure time} \\
\colhead{ } & \colhead{ [au] } & \colhead{[au] } & \colhead{[deg]} & \colhead{[s/image]} }
\decimalcolnumbers
\startdata
2022 Oct 7   & 1.60 & 1.83 & 33.1 & 150 \\
2022 Oct 22  & 1.63 & 1.76 & 34.0 & 150 \\
2022 Nov 1   & 1.66 & 1.71 & 34.3 & 150 \\
2022 Nov 5   & 1.67	& 1.69 & 34.3 & 150 \\
2022 Nov 16  & 1.71	& 1.63 & 34.3 & 150 \\
2022 Nov 25  & 1.75 & 1.59	& 34.0 & 150 \\
2024 Jan 31 & 4.08 & 4.18 & 13.6 & 180 \\
\enddata
\end{deluxetable*}

\begin{figure}[ht!]
\plotone{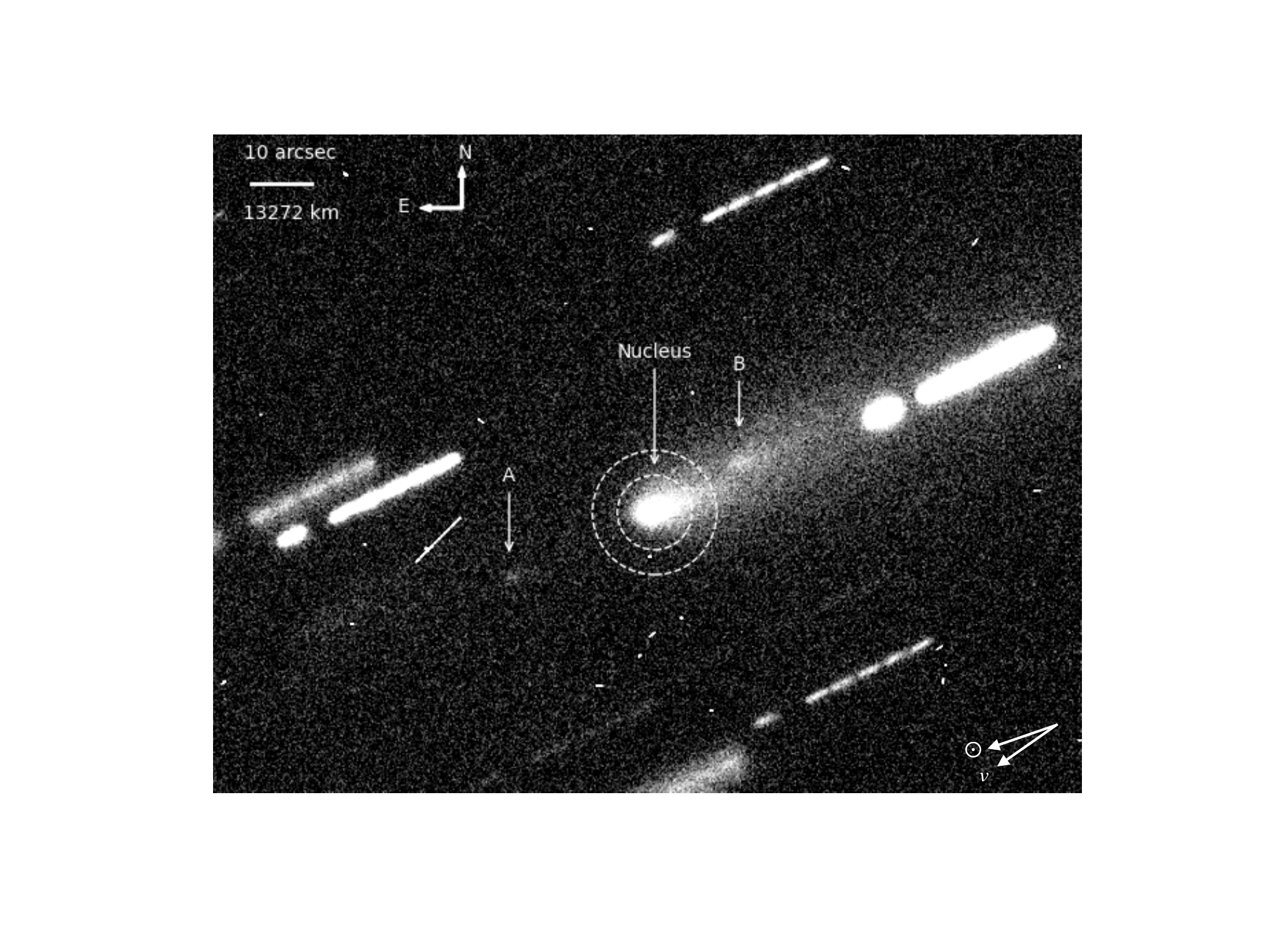}
\caption{A summed image of 157P on UT 2022 October 7, consisting of 6 individual 150s exposures. Fragment A and complex B are indicated with arrows. The oblique trails are due to field stars; the small gap in each trail is due to a short break in the observations.  The dotted circles show the sky annulus.  \label{fig:Fig1}}
\end{figure}

\begin{figure}[ht!]
\plotone{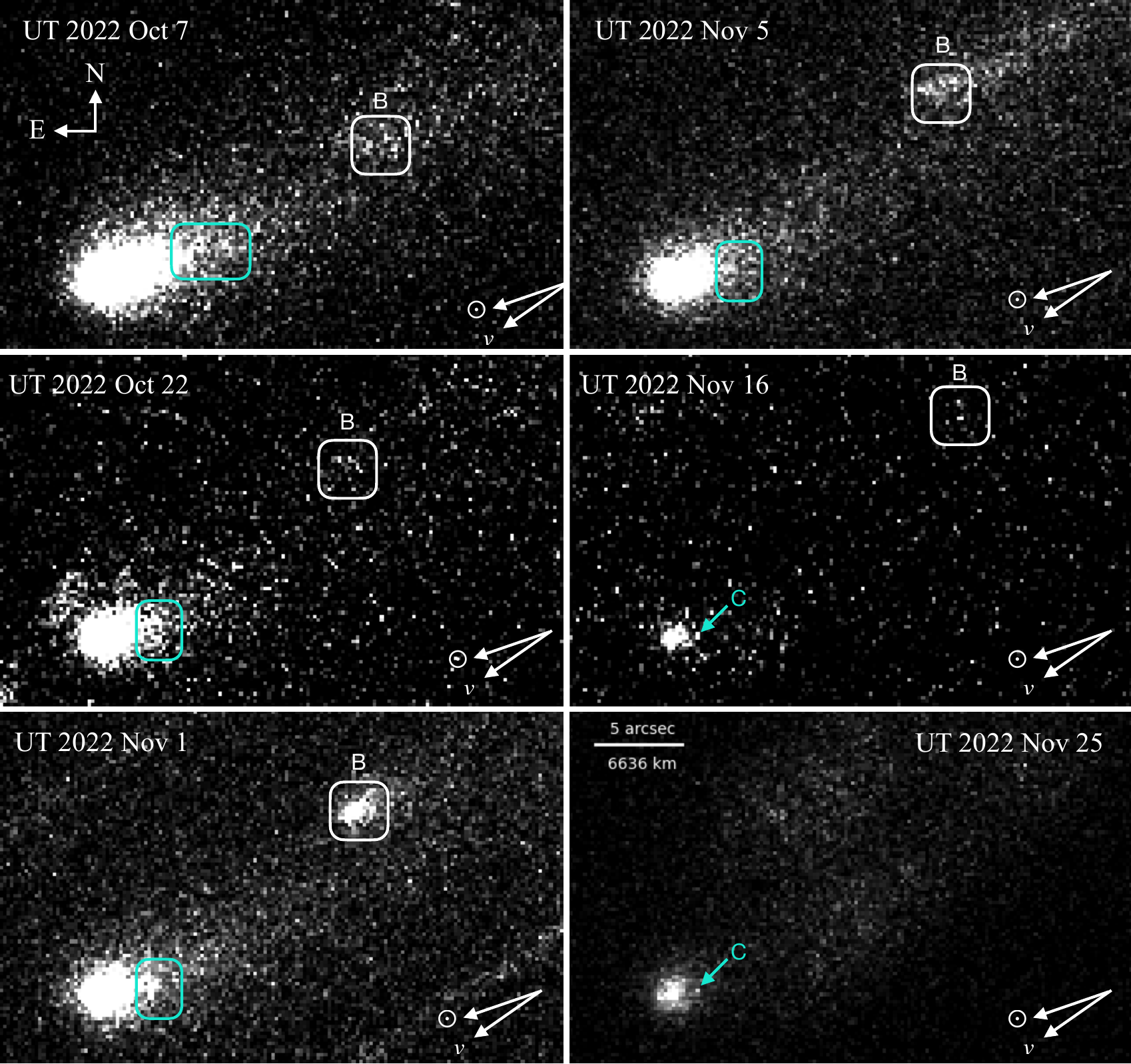}
\caption{Median image from each night of observation.  The green boxes show the near-comet fragments, while the white boxes show the fragment cluster B. Arrows point to Fragment C, visible on UT 2024 November 16 and 25.  All panels have the same dimensions.  \label{fig:Fig2} }
\end{figure}

Each night of observation yielded on average 7 images; the integration time was 150s for the 2022 observations, and 180s for the 2024 observations.  The images were processed by first subtracting a bias, then normalized by an evenly illuminated flat-field image.  The telescope was tracked at 157P’s angular rates, so field stars were slightly trailed in the images.  The geometrical parameters for the observations are provided in Table 1.

\subsection{The comet}

Figure 1 is a summed image of 157P from UT 2022 October 7, created from 6 individual 150s images.  The comet is in the center, the diagonal trails are tracks of field stars and galaxies created by shifting and adding the images.  157P has a moderate coma and a tail pointing toward the northwest, roughly 50\arcsec~ in extent.  In the Figure, a very faint and diffuse blob (``A") can be seen in front (East and South) of the comet, while a bright blob (``B") lies behind the comet (West and North of it).  Component A was detected only on 2022 October 7, its brightness at the limit of our observations, and was not seen again thereafter.  Given the slow speed of the measurable fragments (see Table 3), the simplest, and likeliest, explanation, is that A had simply disintegrated or become too faint to detect.
 Component B turned out to be a cluster of fragments, and its evolution is described further below.

We revisited 157P again with the NOT on UT 2024 January 31, but neither the comet nor any companion was detected in the six individual images taken that night.  A more sensitive summed image of $6 \times 180$s integrations was created by adding the individual images, but this also failed to show the comet.  Based on the non-detection, we were able to estimate an upper limit for the comet's brightness, which will be presented in Section 3.

\subsection{The fragments}

To search for fainter companions, we calculated the median image for each night of observation.  A mosaic of the median images is shown in Figure 2.  

The median images show that, while 157P still sported a strong coma (UT 2022 October 7 - November 5), there were always 1-2 small condensations embedded on the right edge of the coma ($< $ 5\arcsec W of the nucleus, shown in green box in Figure 2); we interpret these condensation to be fragments of the comet.  These fragments changed their location and appearance each night of observation, and given our $\sim$ 2-week revisit time, we were not able to link the individual fragments from one night to the next.  Furthermore, since these fragments were embedded in the coma, we were not able to make meaningful astrometric or photometric measurements of these objects; nevertheless, we want to draw notice to their existence. 

Further away from the comet, blob "B" ($\sim$ 15\arcsec W and $\sim$ 10\arcsec N of the comet, enclosed in white box in Figure 2) turned out to be a cluster of sublimating fragments. The 2022 November 1 image shows that one of these fragments was in outburst. At the next visit on 2022 November 5, the outburst had dissipated, leaving behind a trail of debris along the comet's velocity vector.  By 2022 November 16, little remained of cluster B.  There was no sign of B on 2022 November 25; however, the median image from this night is plagued by traces of imperfectly subtracted field stars, and B could be hidden in the noisy background.  

Figure 2 shows that the detected fragments changed roughly on the same timescale as our $\sim$ 2-week revisit time.  Furthermore, all fragments were clustered near the comet and cluster B, none appeared in the region between the main comet optocenter and cluster B.  Based on these two observations, we deduce the following:
\begin{enumerate}
    \item There were only two sources for the fragments: 157P's nucleus, and a companion located in the vicinity of cluster B. This companion may be one of the fragments in the cluster, or it may have been a completely disrupted precursor object;
    \item The fragments must have lifetimes on the order of $\sim2$ weeks, in order to explain their new appearances at each revisit.  We have no constraint on the reason for their disappearance, although we suspect it is related to spin-up by outgassing torques (see Section 4.1).
    \item Due to their short lifetimes, the fragments simply did not have time to travel far, hence the lack of fragments at greater distances from the two sources. 
\end{enumerate}

Finally, fragment C was the only fragment seen on UT 2022 November 16 and 25, after 157P's coma had largely dissipated. 

\section{Results}

\subsection{Aperture Photometry}

\begin{figure}
\plotone{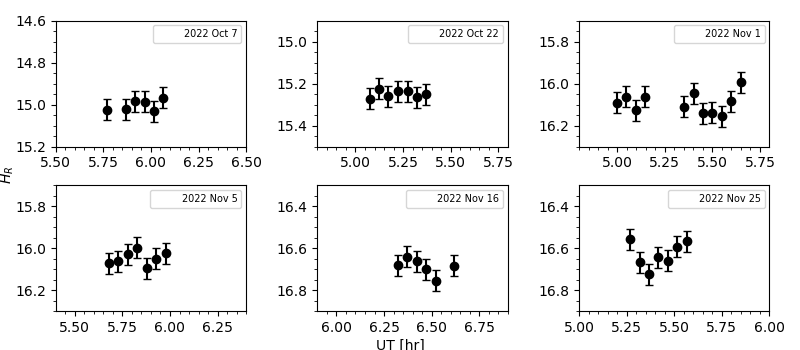}
\caption{157P's nightly photometry. \label{fig:Fig3} }
\end{figure}

\begin{figure}[h!]
\epsscale{0.6}
\plotone{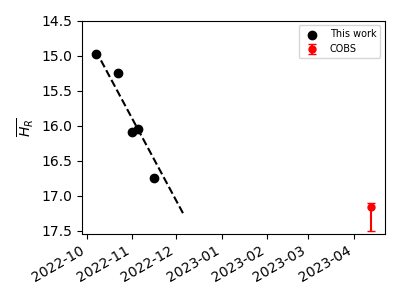}
\caption{157P's nightly mean $H_R$.  The formal uncertainties are $\sim$ 0.03 mag, smaller than the size of markers.  The straight line is a linear fit to our data, with the slope of 0.04 mag/day. The error bar on the COBS point is due to the uncertainty in the filter used. \label{fig:Fig4}}
\end{figure} 

 The measurement of the comet and its fragments was calibrated using differential photometry with nearby field stars that appear in the Pan-Starrs 1 Catalog \citep{tonry2012}.  The seeing was approximately 1\arcsec~(5 pixels Full Width at Half Maximum), but variable.  We tried several photometry apertures, with radii between 7 and 10 pixels, corresponding to 1.5” and 2.1”  respectively.   The 1.5\arcsec~aperture was found to be most consistent from night to night and was used for all the analysis herein, unless stated otherwise. The sky background was calculated from the median intensity inside a concentric annulus with inner radius 6.4\arcsec~and outer radius 10.7\arcsec, then subtracted from the signal.   Experiments show that the median signal in the chosen annulus  provides a good measure of the sky provided the total number of pixels in the annulus is large compared to the number of pixels in the annulus containing emission from the dust tail, as is the case.  We did not use measurements of the sky at larger distances from the comet because these  suffer increasingly from the effects of non-flatness in the data.  For each night of observation, several field stars were selected as reference stars, to make sure the results were consistent.  
 
 For absolute calibration, we made use of all bright field stars that were not close to saturation; each night yielded $\sim 10$ such stars.  The typical mean error on the stellar magnitudes was approximately $\pm$0.01 mag, while  the error on 157P was $\pm$ 0.03 mag. 

To correct for the different viewing geometries over the two-month observation period, we calculate the absolute magnitude $H_R$:

\begin{equation}
H_R = m_R - 5\log(r_H \Delta) - \beta \alpha
\end{equation}

\noindent where $r_H$ and $\Delta$ are the heliocentric and geocentric distances [in au], respectively, $\alpha$ is the phase angle [in degree], and $\beta = 0.04$ mag per degree is the assumed phase darkening coefficient.  Equation (1) is strictly valid only for a point source, or one that is completely contained within the projected photometry aperture.  For a resolved source, the dependence on geocentric distance is slower than $\Delta^2$ but its form depends on the target morphology.  Given that the morphology of 157P is both complex and time dependent, we elect to use the point source approximation and recognize that the resulting error in $H_R$ will be small except for the 2024 January 31 observation, because $\Delta$ varies minimally in all earlier observations.  157P's  photometry from each night is plotted in Figure 3, while Figure 4 shows the nightly mean absolute magnitude $\overline{H_R}$.  Both mean R magnitude $\overline{m_R}$ and $\overline{H_R}$ are also tabulated in Table 2. 

\begin{deluxetable*}{lcccc}
\tablecaption{157P Photometry \label{tab:cometPropTable}}
\tablewidth{0pt}
\tablehead{
\colhead{UT Date} & \colhead{$\overline{m_R}$} & \colhead{$\overline{H_R}$} & \colhead{$C_e$} & \colhead{$r_e$} \\
\colhead{ } & \colhead{ } & \colhead{ } & \colhead{[km$^2$]} & \colhead{[km]}
}
\decimalcolnumbers
\startdata
2022 Oct 7 & 18.67 $\pm$ 0.02 & 14.98 & 24.7 & 2.8 \\
2022 Oct 22	& 18.90 $\pm$ 0.02 & 15.25 & 19.4 & 2.5 \\
2022 Nov 1 & 19.73 $\pm$ 0.03 & 16.09 & 8.9 & 1.7 \\
2022 Nov 5 & 19.67 $\pm$ 0.02 & 16.05 & 9.3 & 1.7 \\
2022 Nov 16	& 20.35 $\pm$ 0.06 & 16.75 & 4.9 & 1.2 \\
2022 Nov 25	& 20.19 $\pm$ 0.05 & 16.63 & 5.4 & 1.3 \\
2024 Jan 31 & $> 23.8$ & $> 17.1$ & $< 3.5$ & $< 1.1$ \\
\enddata
\end{deluxetable*}

The cross section of the comet, $C_e$ (in km$^2$), contributed by both nucleus and dust, can be estimated from $H_R$: 

\begin{equation}
C_e = \frac{\left(2.25 \times 10^{16}  \right) \pi }{p_R} 10^{\left[-0.4 \left(H_R -  m_{\odot} \right) \right]},
\end{equation}

\noindent where $p_R = 0.04$ is the assumed geometric albedo in the R filter, and $m_{\odot} = -27.15$ is the apparent R magnitude of the Sun \citep{willmer2018}. The  radius of a circle of equal area is
then $r_e = \sqrt{C_e/\pi}$.  Both $C_e$ and $r_e$ are also listed in Table 2.  Note that coma contamination has not been accounted for, so the values presented here represent upper limits to the cross-section and radius of the nucleus. 

The Comet Observation database (COBS) reported an observation for 157P on UT 2023 April 13 (https://cobs.si/obs/comet/105/), when the comet was at $r_H = 2.54$ au, $\Delta = 1.74$ au, and phase angle $16.57^o$.  The reported magnitude was 21.4, with no filter information.  Assuming the measurement was made in the V filter, and assuming solar colors V-R = 0.35 \citep{willmer2018}, we calculate a corresponding absolute magnitude in the R filter $H_R = 17.16$, cross section $C_e = 3.3$ km$^2$, and effective radius $r_e = 1.0$ km.  The COBS data point is also plotted alongside our data in Figure 4, where the error bar allows for the possibility that the original COBS data might be in the R filter.  If this data point is reliable, it captured the comet at an even fainter state than our late 2022 data.  The evolution of $C_e$ with time will be discussed further in Section 4.2.

Fitting our data in Figure 4 indicates that 157P faded at a rate of 0.04 mag/day during our observing campaign. The coma from UT 2022 October 7 dwindled quickly to being barely visible by late 2022 November.  The effect of the fading coma can be seen in Figure 3: as the coma faded, the nucleus's contribution to the photometry became more dominant and indeed, the comet brightness on UT 2022 November 16 and 25 shows cyclic variations about a mean $H_R$ = 16.7 mag, with a range of $\sim $ 0.2 mag, suggestive of a rotating nucleus. While the data are not sufficient to determine an unambiguous rotation period,  they are suggestive of a fast spin, perhaps with a spin period $\sim$ a few hours. 

\begin{deluxetable*}{lccccc}
\tablecaption{Fragment Separation \label{tab:fragdyntable}}
\tablewidth{0pt}
\tablehead{
\colhead{UT Date} & \colhead{Fragment} & \colhead{$\Delta$RA} &
\colhead{$\Delta$Dec}& \colhead{Separation} & \colhead{Separation velocity} \\
\colhead{ } & \colhead{ } & \colhead{[arcsec]} &
\colhead{[arcsec] } & \colhead{[$\times 10^3$ km] } & \colhead{[m/s]}
}
\decimalcolnumbers
\startdata
2022 Oct 7 & A$^1$ & 17.8 & 23.6 & 33.4 &  -- \\
           & B$^2$	& -15.24 & 8.14 & 23.6 & 1.3 \\
           &  &  &  & & \\
2022 Oct 22 & B$^2$ & -14.54 & 10.70 & 23.5 & 1.3 \\
         &  &  &  & & \\
2022 Nov 1 & B$^2$ & -14.74 & 10.94 & 23.1 & 1.3 \\
         &  &  &  & & \\
2022 Nov 5 & B$^2$ & -16.08 & 11.98 & 24.5 & 3.1 \\
         &  &  &  & & \\
2022 Nov 16	& B$^2$ & -16.14 & 12.88 & 27.2 & 3.1 \\
            & C & -1.46 & 0.04 & 1.79 & 0.3 \\
         &  &  &  & &\\
2022 Nov 25	& C & -1.52 & 0.23 & 1.88 & 0.3 \\
\enddata
\tablecomments{\  $^1$ Fragment A was detected only on one night, too short to measure a velocity.\\
$^2$ B is not a single object, but a cluster of fragments whose appearance changed over time.  The separations listed here refer to the brightest object in the cluster, not any particular fragment.}
\end{deluxetable*}


\subsection{Fragment Properties}
The fragments of 157P were often faint, so their positions were measured from the nightly stacked images in order to minimize errors.  The positions were determined by centroiding the fragments within a 5-pixel wide box;  if the centroiding was clearly wrong due to confusion with a nearby bright star, the centroiding was done by eye.  We estimate that the positional uncertainty was $\sim$ 1 pixel (0.214\arcsec). 

\begin{figure}
\plotone{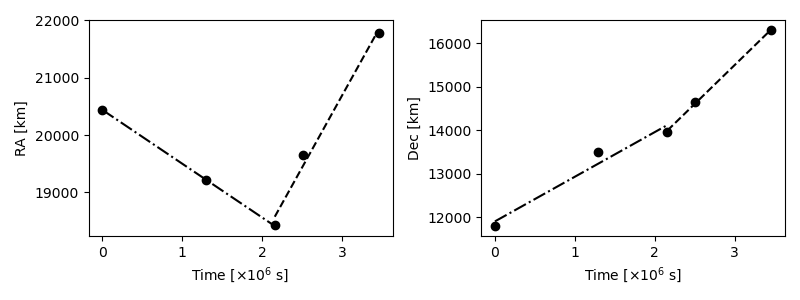}
\caption{Motion of the brightest pixel in fragment cluster B as a function of time, in RA (left) and Dec (right).  The x-axis is the time elapsed since the time of the first observation at UT 2022 October 7 05:46:09.  The cluster changes direction and magnitude after UT November 1.  The lines shown are the fit to the motion over UT October 7 - November 1 ($dRA/dt = -0.9$ m/s, $d(Dec)/dt = 1.0$ m/s), and UT November 1 - 16 ($dRA/dt = 2.5$ m/s, $d(Dec)/dt = 1.8$ m/s).  \label{fig:Fig5} }
\end{figure}

\begin{deluxetable*}{lcllcc}
\tablecaption{Fragment Photometry \label{tab:fragphottable}}
\tablewidth{0pt}
\tablehead{
\colhead{UT Date} & \colhead{Fragment} & \colhead{$\overline{m_R}$} &
\colhead{$\overline{H_R}$} & \colhead{$C_e$} & \colhead{$r_e$} \\
\colhead{ } & \colhead{ } & \colhead{ } &
\colhead{ } & \colhead{[km$^2$]} & \colhead{[km]}}
\decimalcolnumbers
\startdata
2022 Oct 7 & A	& 22.03 $\pm$ 0.10 & 18.35 $\pm$ 0.10 & 1.1 & 0.6 \\
           & B & 20.98 $\pm$ 0.05 & 17.33 $\pm$ 0.05 & 2.9 & 1.0 \\
         &  &  &  &  & \\
2022 Oct 22	& B	& 21.42 $\pm$ 0.08 & 17.78 $\pm$ 0.08 & 1.9 & 0.8 \\
         &  &  &  &  &  \\
2022 Nov 1 & B & 21.21 $\pm$ 0.03 & 17.58 $\pm$ 0.03  & 2.3 & 0.8 \\
         &  &  &  &  & \\
2022 Nov 5 & B & 21.46 $\pm$ 0.04 & 17.84 $\pm$ 0.04  & 1.8 & 0.8 \\
         &  &  &  &  & \\
2022 Nov 16	& B	& 22.31 $\pm$ 0.12 & 18.71 $\pm$ 0.12 & 0.8 & 0.5 \\
         & C & 20.59 $\pm$ 0.03   & 17.00 $\pm$ 0.03   & 3.9 & 1.1 \\
         &  &  &  &  & \\
2022 Nov 25	& C	& 20.73 $\pm$ 0.03 & 17.15 $\pm$ 0.03 & 3.4 & 1.0 \\
\enddata
\end{deluxetable*}

The fragment separations from the comet are summarized in Table 3.  Fragment A was observed only on one night so no separation velocity was available. Given the small velocities for all detected fragments, it was unlikely that fragment A had moved out of our field of view after the initial observations.  The simplest, and likeliest, explanation for the non-detection of fragment A after UT 2022 October 7 is that it had disintegrated or become too faint for detection, like many of the other fragments.

Cluster B was far enough away from the comet that photometry and astrometry was possible. For UT 2022 October 7 - November 16, we measured the position of the brightest object in B; we caution that the brightest object most likely does not refer to the same object each night.  The RA and Dec motion of cluster B is plotted as a function of time in Figure 5.  The Figure shows that B's motion changes in both direction and magnitude after UT 2022 November 1.  Fitting the RA and Dec motion separately, we measure $dRA/dt = -0.9$ m/s for UT October 22 - November 1 and $2.5$ m/s for UT November 1 - 16; similarly, $d(Dec)/dt = 1.0$ m/s and $1.8$ m/s for the same time periods.  It is not clear whether this break in motion is due to the changing observational geometry, or reflects misidentification of the fragments on different dates.

The most reliable separation velocity is that of fragment C, which was the only fragment whose identity was secure and was observed on two nights.  C's separation velocity was very small, $\sim 0.3$ m/s.  The implication of this is discussed in Section 4.1.


Table 4 lists the mean R magnitude, mean absolute magnitude, cross-sections and effective radii for fragments A, B, and C.  Fragment A was too faint for reliable measurement in single exposures, so we calculated its brightness from the UT October 7 stacked image. B was visible until UT 2022 November 25, when we failed to detect it.  However, the images from that night were also degraded by the presence of many bright field stars, so we suspect that the non-detection of B may be an artifact of the poor image quality. 

The Minor Planet Center published an orbit for a companion called "157P-B," but \citet{sekanina2023} proposed that the 157P-B's astrometry was best explained by not one, but two separate companions, which he called "C" and "D"."  Sekanina's C was located West and North of the comet -- in the same direction as our cluster B \ -- while his D was West and South. It is tempting to identify our cluster B with Sekanina's C, but there are not enough data to clearly link Sekanina's C and D with any fragment reported in this work. We note, however, that Sekanina's hypothesis of a fragment's replacement by another similar in brightness is very much in keeping with our observations. As Figure 2 shows, the detected fragments changed roughly on the same timescale as our revisit time, suggesting that most fragments had lifetimes on the order of $\sim$ 2 weeks.

\section{Discussion}
\subsection{Rotational instability as cause of splitting}

During most of our campaign, 157P displayed a significant coma and, accordingly, the photometry from most nights shows no periodic modulation typical of a rotating nucleus.  However, the photometry on 2022 November 16 and 25, does show  systematic brightening and fading that suggests a short rotation period on the order of 2 - 3 hr.  Such fast rotation is not unprecedented among comets. The fastest spinning comet known is the SOHO comet 322P, with a 2.8 $\pm 0.3$ hr rotation period \citep{knight2016}.  The critical rotation period for rotational instability in a spherical fluid (i.e., strengthless) body is:

\begin{equation}
P_c = \sqrt{\frac{3 \pi}{G\rho}},    
\end{equation}

\noindent where 
$G = 6.67\times10^{-11}$ N kg$^{-2}$ m$^2$
is the gravitational constant. A strengthless body spinning faster than $P_c$ will have its maximum centrifugal force exceed its gravitational force and thus becomes susceptible to mass shedding.  In this simple model, $P_c$ depends only on the bulk density, $\rho$, and 
substituting $\rho = 600$ kg/m$^3$ yields $P_c = 4.3$ hr. If 157P spins at a rate of 2 - 3 hr, as suggested by Figure 3, rotational instability may be at the root of its splitting.   The strongest limit on the nucleus radius derived from our photometry is $\leq 1.1$ km (Table 2);  with $\rho = 600$ kg/m$^3$, this corresponds to an escape velocity $v_{esc} \sim$ 0.6 m/s.  The fragment  separation velocities are $\lesssim 1$ m/s, comparable to $v_{esc}$, and thus consistent with mass shed from the nucleus by centrifugal forces.  

We note that such rotationally-induced mass shedding has been observed in several asteroids (e.g., Gault, \citet{luu2021}), and comets (\citet{jewitt2021},\citet{jewitt2022}).  In a direct analogy with 157P, the splitting of comet 332P/Ikeya-Murakami, which rotates with period $\sim$ 2 - 3 hr, is also believed to be due to its fast rotation \citep{jewitt2016}. 

Two mechanisms are capable of changing the spin rate of small bodies in the solar system: the YORP torque (due to the momentum carried by photons) and outgassing torque (due to the sublimation of near surface ice).  YORP spin-up times for bodies of kilometer scale are measured in millions of years, while sublimation spin-up times are $\sim$ 4 - 5 orders of magnitude shorter, all else being equal.  For a comet with radius $r$, and perihelion in the 1 to 2 au distance range,  an empirical estimate of the spin-up timescale due to outgassing torque is \cite{jewitt2021}:

\begin{equation}
\tau_{outgas} \ {\rm{[yr]}} \sim 100 \ \left(\frac{r}{1 \ {\rm km}} \right)^2    
\end{equation}

According to Equation (4), with $r_e =$ 1 km, outgassing torque can double 157P's spin in just 100 yrs, a tiny fraction of the $\sim$0.5 Myr dynamical lifetime of short-period comets \citep{levison&duncan1993}.  The same equation implies that the $\sim$ 2-week lifetime we found for the fragments (Section 2.2) corresponds to a fragment radius $\sim$ 20 m.  This is much smaller than the  radii in Table 4, and it might appear that outgassing torques could not be responsible for the fragments' removal.  However, the radii in Table 4 are strong upper limits to the true radii, since all the fragments were sublimating and their effective radii refer mostly to the total cross section in dust.  The bare fragments may well be $\sim$ 20 m in radius. 


For 157P, the YORP torque is $\sim 10^5$ times weaker than the outgassing torque and therefore unlikely to play an important role here.

\subsection{The fading of 157P}

\begin{figure}
\epsscale{0.5}
\plotone{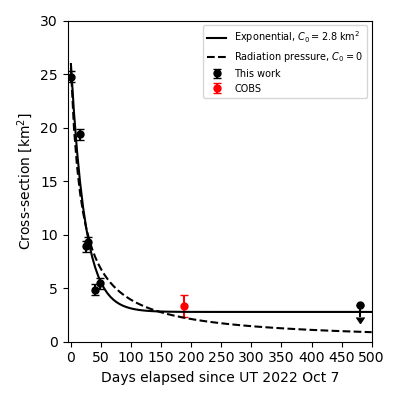}
\caption{Scattering cross-section of 157P over time.  The lines are the fits to the data using Eqs. (7) and (8). \label{fig:Fig6} }
\end{figure}

Figure 6 shows the decline of 157P's cross-section over time.  The comet faded sharply over the first few nights of observations, then seemed to reach an asymptote. Here we consider two possible explanations for 157P's fading: (1) the comet faded due to coma dust particles being swept out of the photometry aperture by radiation pressure, and (2) the comet was shrinking due to continuous fragmentation.

\subsubsection{Radiation pressure sweeping}

The first scenario assumes that dust particles are released instantaneously from the nucleus at zero velocity, then accelerated by radiation pressure out of the photometry aperture.  Radiation pressure sweeps a particle of radius $a$ a distance $\ell$ on timescale (Jewitt 2021):

\begin{equation}
\tau_{rad} \sim \left[\frac{2\ell}{\beta_{r} g_{\odot}(r_H)}\right]^{1/2},
\end{equation}

\noindent where $\beta_r$ is the dimensionless radiation pressure factor, so that $\beta_r g_{\odot}(r_H)$ is the particle acceleration.  Using 1.7 au as the representative heliocentric distance, $g_{\odot}(1.7) = 2 \times 10^{-3}$ m/s$^{-2}$. $\beta$ depends on the wavelength and many unknown parameters like the composition and size of the particle, but it can be approximated as $\beta \sim 10^{-6}/a$ \citep{bohren&huffman1983}, where $a$ is in meters.  With the photometry radius $\ell = 1.8 \times 10^6$ m, the representative geocentric distance also 1.7 au, $\tau_{rad} \sim 490 a^{1/2}$ days.  Setting $\tau_{rad}$ equal to our shortest revisit time, 4 days, we find that radiation pressure can explain the comet's changing appearance if the particle radius is $a \le 67 \ \micron$.

The dust size distribution is usually assumed to obey a power law size distribution:

\begin{equation}
n(a) da = \Gamma a^{-q} da,
\end{equation}

\noindent where $n(a) da$ is the number of particles with radii between $a$ and $a + da$, and $\Gamma$ and $q$ are constants. With the nominal value $q = 3.5$ typical of dust released from asteroids and comets, the fading of a comet due to radiation pressure can be described by \citet{jewitt2017}:

\begin{equation}
C(t) = C_0 + \frac{K}{t - T_0}.
\end{equation}

\noindent where $C(t)$ is the comet's total scattering cross-section, $C_0$ is the contribution from the nucleus, and the second term the contribution from the coma.   $K$ and $T_0$ are constants; $K$ is related to the size of the photometry aperture and $T_0$ the time the particles were released from the nucleus.  Fitting the data to Equation (7) yields the best-fit parameters are $C_0 = 0, K = 462.1$, and $T_0 = -18.1$; the fact that $T_0 < 0$ is not significant since it is simply due to the fact that we have chosen $t = 0$ to be UT 2022 October 7.  The fit is also shown in Figure 6. This model implies that 157P's cross-section is controlled by the coma, with little or no contribution from the nucleus, and that the comet fades simply because dust particles are removed from the photometry radius.   

\subsubsection{Continuous fragmentation}

In the second scenario, the comet's shrinking cross-section is attributed to a nucleus that fragments repeatedly, at an average rate that is proportional to the remaining nucleus size.  Such repeated mass shedding events is expected if the rotation rate remains high enough to sustain large centrifugal forces.  Repeated fragmentation would give rise to an exponential decline in the comet's cross-section: 

\begin{equation}
C(t) = C_0 + A \exp^{-B \Delta t},
\end{equation}

\noindent where $C_0$ is again the contribution from the nucleus, $A$ is a scaling constant, $B$ is the exponential decay rate, and $\Delta t$ is the time elapsed (in days) since our first observation on UT 2022 October 7.  Fitting the data in Figure 6 to Equation (8) yields the best-fit values $C_0 = 2.8$ km$^2$, $A = 23.2$  and $B = 0.04$/day; this fit is also plotted in Figure 6. This best-fit $C_0$ implies the effective radius $\sqrt{C_0/\pi} = 0.9$ km.  This is consistent with $r_e = 1.0$ km from the 2023 data point, as it should be.  The most important parameter in this model is $B$, which tells us that the exponential decay time is $B^{-1} \sim$ 25 days. The comet's cross-section shrinks at the rate of, on average, 4\% per day.

Figure 6 shows that both Equations (7) and (8) provide plausible fits to the  fading of 157P, but with exponential decay providing the better fit to the late stage photometry.  We cannot exclude the possibility of other models.  For example, the fading could be due to radiation pressure sweeping of particles released over a finite period, instead of impulsively, as assumed.  It could also be due to radiation pressure and exponential decay combined.  By Occam's razor, however, we proceed on the assumption that exponential decay is the operative process.

Interestingly, 157P has already gone through a similar cycle of outburst and fragmentation on the previous orbit.  Based on observations from late 2016 and early 2017, \citet{sekanina2023} postulated that 157P already fragmented in early 2017, with the fragmenting event accompanied by an outburst.  We observed similar outburst-fragmentation events in both the nucleus and the fragments of cluster B (see Figure 2).  The relationship between outburst and fragmentation is not clear, but may be related to outgassing torques (from the outburst) spinning up the nucleus or a fragment to the point of mass shedding (Equation 4).  Note that for small radii, Equation 4 indicates that the sublimating object will enter a feedback loop where a small mass leads to a short spin-up timescale, resulting in mass loss, which then results in an even shorter spin-up timescale and more mass loss, etc.  There is no obvious mechanism to stop this runaway process other than the intrinsic cohesive strength of the nucleus material.

We recognize that the two models presented above are not unique, and that other models might fit the data just as well. These models are useful because they represent two extreme ends of what might explain a comet's fading. The repeated fragmentation model asserts that the comet fades because the coma dust source keeps fragmenting.  In contrast, the radiation sweeping model assumes that the comet fades because dust grains impulsively released into the coma are swept out of the photometry aperture by radiation pressure.  In reality, these mechanisms are not exclusive of each other, and both are likely to play a role in cometary fading.  Future observations may be able to break the degeneracy, if the observations are sensitive enough to either detect the nucleus, or impose more stringent upper limits on the nucleus size.



\clearpage
\section{Summary}

We monitored comet 157P/Tritton in October and November 2022, with additional observations on UT 2024 January 31. The main results of our observations are:

\begin{itemize}
\item 157P was highly active in early October 2022 but faded dramatically by late November, despite only minimal changes in the observing geometry.  In its later, low activity state, 157P showed cyclic brightness variations that suggested a fast rotation period of a few hours, although a precise period cannot be identified in the available data (Figure 3).

\item  Several fragments were detected, either very near the nucleus, or in a cluster $\sim 20$\arcsec \ West and North of the comet (Figure 2). The fragments changed appearance at every revisit, suggesting repeated fragmentation, and fragment lifetimes were comparable to the revisit time ($\sim 2$ weeks). 

\item We measured only upper limits to the sizes of the 157P nucleus and its fragments, because of strong contamination by ejected dust.  With a nucleus radius $\leq 1.1$ km and assumed bulk density 600 kg/m$^3$, the escape velocity is $v_{esc} \sim$ 0.6 m/s.  The fragment  separation velocities are $\lesssim 1$ m/s, comparable to $v_{esc}$, and consistent with mass loss by centrifugal forces.

\item The comet's decreasing cross-section is most simply fitted  with  a fragmentation model having exponential decay timescale $B^{-1}$ = 25 days.

\item The difficulty in correlating fragments from epoch to epoch highlights the need for better temporal monitoring of fragmenting comets.
\end{itemize}

\begin{acknowledgments}
We thank the referee for thoughtful and helpful comments.  We are grateful to Anlaug Amanda Djupvik, John Telting, and all the NOT observers for help with the NOT observations.  We also thank the Comet OBServation database (COBS) for their service to comet observers.\\

\noindent The Research Council of Norway (RCN), through its Centers of Excellence funding scheme, project number 332523 (PHAB), is acknowledged for financial support.

\end{acknowledgments}

%

\vspace{5mm}
\facilities{Nordic Optical Telescope}






\bibliography{paper157P}{}

\begin{thebibliography}{}
\expandafter\ifx\csname natexlab\endcsname\relax\def\natexlab#1{#1}\fi
\providecommand{\url}[1]{\href{#1}{#1}}
\providecommand{\dodoi}[1]{doi:~\href{http://doi.org/#1}{\nolinkurl{#1}}}
\providecommand{\doeprint}[1]{\href{http://ascl.net/#1}{\nolinkurl{http://ascl.net/#1}}}
\providecommand{\doarXiv}[1]{\href{https://arxiv.org/abs/#1}{\nolinkurl{https://arxiv.org/abs/#1}}}

\bibitem[{{Bohren} \& {Huffman}(1983)}]{bohren&huffman1983}
{Bohren}, C.~F., \& {Huffman}, D.~R. 1983, {Absorption and scattering of light
  by small particles} (Wiley, New York)

\bibitem[{{Jewitt}(2021)}]{jewitt2021}
{Jewitt}, D. 2021, \aj, 161, 261, \dodoi{10.3847/1538-3881/abf09c}

\bibitem[{{Jewitt}(2022)}]{jewitt2022}
---. 2022, \aj, 164, 158, \dodoi{10.3847/1538-3881/ac886d}

\bibitem[{{Jewitt} {et~al.}(2017){Jewitt}, {Agarwal}, {Li}, {Weaver},
  {Mutchler}, \& {Larson}}]{jewitt2017}
{Jewitt}, D., {Agarwal}, J., {Li}, J., {et~al.} 2017, \aj, 153, 223,
  \dodoi{10.3847/1538-3881/aa6a57}

\bibitem[{{Jewitt} {et~al.}(2016){Jewitt}, {Mutchler}, {Weaver}, {Hui},
  {Agarwal}, {Ishiguro}, {Kleyna}, {Li}, {Meech}, {Micheli}, {Wainscoat}, \&
  {Weryk}}]{jewitt2016}
{Jewitt}, D., {Mutchler}, M., {Weaver}, H., {et~al.} 2016, \apjl, 829, L8,
  \dodoi{10.3847/2041-8205/829/1/L8}

\bibitem[{{Knight} {et~al.}(2016){Knight}, {Fitzsimmons}, {Kelley}, \&
  {Snodgrass}}]{knight2016}
{Knight}, M.~M., {Fitzsimmons}, A., {Kelley}, M. S.~P., \& {Snodgrass}, C.
  2016, \apjl, 823, L6, \dodoi{10.3847/2041-8205/823/1/L6}

\bibitem[{{Levison} \& {Duncan}(1993)}]{levison&duncan1993}
{Levison}, H.~F., \& {Duncan}, M.~J. 1993, {The long-term dynamical behavior of
  short-period comets}, In its Origins of Solar Systems Workshop: The Origin,
  Evolution, and Detectability of Short Period Comets p. 1-36 (SEE N94-11628
  01-90)

\bibitem[{{Luu} {et~al.}(2021){Luu}, {Jewitt}, {Mutchler}, {Agarwal}, {Kim},
  {Li}, \& {Weaver}}]{luu2021}
{Luu}, J.~X., {Jewitt}, D.~C., {Mutchler}, M., {et~al.} 2021, \apjl, 910, L27,
  \dodoi{10.3847/2041-8213/abedbc}

\bibitem[{{Sekanina}(2023)}]{sekanina2023}
{Sekanina}, Z. 2023, arXiv e-prints, arXiv:2309.01923,
  \dodoi{10.48550/arXiv.2309.01923}

\bibitem[{{Tonry} {et~al.}(2012){Tonry}, {Stubbs}, {Lykke}, {Doherty},
  {Shivvers}, {Burgett}, {Chambers}, {Hodapp}, {Kaiser}, {Kudritzki},
  {Magnier}, {Morgan}, {Price}, \& {Wainscoat}}]{tonry2012}
{Tonry}, J.~L., {Stubbs}, C.~W., {Lykke}, K.~R., {et~al.} 2012, \apj, 750, 99,
  \dodoi{10.1088/0004-637X/750/2/99}

\bibitem[{{Willmer}(2018)}]{willmer2018}
{Willmer}, C. N.~A. 2018, \apjs, 236, 47, \dodoi{10.3847/1538-4365/aabfdf}

\end{thebibliography}
\bibliographystyle{aasjournal}



\end{document}